\documentclass[preprint,AIP]{revtex4}
\usepackage{amsmath,amssymb}

\begin{document}

\author{Hironobu Kihara}

\affiliation{
3-26-3-104, Minami-Senzoku, Ota, Tokyo 145-0063, Japan
}

\title{Equations of Motion Solved by the Cremmer-Scherk Configuration on Even-Dimensional Spheres}

\begin{abstract}
Equations of motion of low-energy effective theories of quantum electrodynamics include infinitely many interaction terms, which make them difficult to solve.
The self-duality property has facilitated research on the solutions to these equations. 
In this paper, equations of motion of systems of non-Abelian gauge fields on even-dimensional spheres are considered. 
It is demonstrated that the Cremmer-Scherk configuration, 
which satisfies certain generalized self-duality equations, 
becomes the classical solution for the class of systems that are given 
by arbitrary functions of class $C^1$ of $2m+1$ quantities. 
For instance, Lagrangians consisting of multi-trace terms are included in this class.
This result is likely to generate several new and interesting directions of research, including the classification of actions with respect to the stability condition against the Cremmer-Scherk configuration.
\end{abstract}

\maketitle

Low-energy effective theories of quantum electrodynamics include higher order terms of field strengths. 
In general, it is difficult to solve equations of motion in these theories. 
Self-duality equations and generalized self-duality equations simplify the problems in some cases. 
Cremmer and Scherk studied grand unified theories in higher dimensional space-times. They considered gauge configurations on spheres in the context of spontaneous compactification \cite{Cremmer:1976ir}, and their configurations satisfy the generalized self-duality equations derived in an independent study \cite{Tchrakian:1978sf}. 
The self-duality property establishes the stability of Cremmer and Scherk configurations at least in the models with certain quartic terms in the field strengths, which were introduced by Tchrakian. 
Further consideration of generalized self-duality on spheres was done. 
Tchrakian {\it et al.}\cite{Tchrakian:1978sf,Bais:1985ns} compared the Hodge dual of some monomials of a field strength two-form, $*F^{\wedge p}$, with other monomials, $F^{\wedge q}$. 
Kihara subsequently extended this scheme to polynomials \cite{Kihara:2011jc}.

In this article, equations of motion derived from pseudo-energies given by arbitrary functions of class $C^1$ are considered. 
The present paper demonstrates that even though an action includes infinitely many interaction terms, 
its equations of motion may be solved by the configuration of Cremmer and Scherk.

Every discussion in this article considers a $2m$-dimensional sphere $S^{2m}$ whose radius is $R$.
An atlas of the sphere is parameterized by a set of $2m$ real parameters, $(\zeta^1 , \cdots , \zeta^{2m})$.

The sphere is endowed with the round metric $ds^2= { | d \zeta |^2}/{(1+ |\zeta |^2/4R^2 )^2}$, and the vielbeins are denoted as $V^i := {d \zeta^i}/{(1+ |\zeta |^2/4R^2 )}$. The volume form is $dv= V^1 \wedge V^2 \wedge  \cdots \wedge V^{2m}$. The definition of the Hodge dual operator acting on the bases of differential forms is given as $* ( V^{a_1} \wedge \cdots \wedge V^{a_p} ) = (1/(2m-p)!)\epsilon^{a_1, \cdots , a_p, a_{p+1}, \cdots , a_{2m}} V^{a_{p+1}} \wedge \cdots \wedge V^{a_{2m}}$. Here, the Einstein summation rule is used. The Hodge dual operator, $*$, linearly acts on the space of differential forms.

Suppose that the gauge group is SO($2m$).
The gauge field one-form, $A$, on a $2m$-dimensional sphere is defined as $A:= (1/2) A_{\mu}^{a,b} \gamma_{a,b} d \zeta^{\mu}$. Here,  $\gamma_a$ $(a=1,2, \cdots, 2m)$ are generators of a Clifford algebra, $\{ \gamma_a , \gamma_b \} = 2 \delta_{a,b}$, and $\gamma_{a,b}= (1/2)[\gamma_a, \gamma_b]$. $\gamma$s are represented on $2^m$-dimensional space.
The chiral matrix is defined as $\gamma_{2m+1}:=i^{-m} \gamma_1 \gamma_2 \cdots \gamma_{2m}$. The square of the chiral matrix is the unit matrix $\gamma_{2m+1}^2={\bf 1}_{2^m}$.
Antisymmetric products of $\gamma$ matrices are written in the following form: $\displaystyle \gamma_{a(1),a(2), \cdots, a(p)} :=\frac{1}{p!} \sum_{\sigma \in {\mathfrak S}_p} {\rm sgn}(\sigma) \gamma_{a(\sigma(1))}  \gamma_{a(\sigma(2))}  \cdots \gamma_{a(\sigma(p))} ~,~~~(a(i)=1,\cdots, 2m , i=1,\cdots, p) $.
The field strength two-form is defined as $F:= dA + q A \wedge A$, where $q$ is the gauge coupling constant.

Quantities $\sigma_p$, $\tau_p$, and $\rho_m$~$(p=1,2, \cdots , m)$ are defined as follows:
\begin{align}
\sigma_{p} &:= *\left[ (-1)^p {\rm Tr} \left\{ F^{\wedge p} \wedge * F^{\wedge p} \right\} \right] ~~,\\
\tau_{p} &:=  * \left[  (-1)^p \left\{ {\rm Tr} F^{\wedge p} \right\}
\wedge *  \left\{ {\rm Tr} F^{\wedge p} \right\} \right]  ~~,\\
\rho_m &:= *{\rm Tr} F^{\wedge m}~~.
\end{align}
Suppose that ${\cal F}(X_1, \cdots ,X_m , Y_1 , \cdots , Y_m, Z)$ is a function of class $C^1$.
The equation of motion obtained from the following pseudo-energy is the focus of this article.
\begin{align}
{\cal E}[A] &:= \int_{S^{2m}} dv {\cal F} (\sigma_1 , \cdots , \sigma_m , \tau_1 , \cdots , \tau_m ,   \rho_m)  ~.
\label{eqn:genaction}
\end{align}
Varying the gauge fields $A \rightarrow A' = A + \delta A$ yields the variation of the pseudo-energy:
\begin{align}
\delta {\cal E} &= \int_{S^{2m}} dv \left\{ \sum_{p=1}^m \frac{\partial {\cal F}}{\partial X_p}
(\sigma_1 , \cdots , \sigma_m , \tau_1 , \cdots , \tau_m ,  \rho_m) \delta \sigma_p \right. \cr
&+ \sum_{p=1}^m \frac{\partial {\cal F}}{\partial Y_p}
(\sigma_1 , \cdots , \sigma_m , \tau_1 , \cdots , \tau_m ,  \rho_m) \delta \tau_p  \cr
& \left. + \frac{\partial {\cal F}}{\partial Z}
(\sigma_1 , \cdots , \sigma_m , \tau_1 , \cdots , \tau_m ,  \rho_m) \delta \rho_m  \right\} ~~,
\end{align}
where the variations of $\sigma_p$, $\tau_p$, and $\rho_m$ are given as follows:
\begin{align}
\delta \sigma_p &= * \left[ 2 (-1)^p \sum_{i=0}^{p-1} {\rm Tr} \left\{  F^{\wedge i} \wedge D (\delta A) \wedge F^{\wedge p-i-1} \wedge * F^{\wedge p} \right\} \right] ~~,\\
\delta \tau_{p} &:=  * \left[  2(-1)^p \sum_{i=0}^{p-1} \left\{ {\rm Tr} F^{\wedge i} \wedge D(\delta A) \wedge F^{\wedge p-i-1} \right\}
\wedge *  \left\{ {\rm Tr} F^{\wedge p} \right\} \right]~~,\\
\delta \rho_{m} &:= \sum_{i=0}^{m-1} {\rm Tr} F^{\wedge i} \wedge D(\delta A) \wedge F^{\wedge m-i-1} \cr
&= m  {\rm Tr} D(\delta A) \wedge F^{\wedge m-1}~~.
\end{align}
Because $*dv=1$ and $\partial S^{2m} = \phi$, the variation can be simplified as
\begin{align}
\delta {\cal E} &= - \sum_{p=1}^m \sum_{i=0}^{p-1} 2 (-1)^p {\rm Tr} \int_{S^{2m}}  D \left\{  \frac{\partial {\cal F}}{\partial X_p}   \left[ F^{\wedge p-i-1} \wedge \left( * F^{\wedge p} \right) \wedge  F^{\wedge i}  \right]  \right. \cr
&+ \left.  \frac{\partial {\cal F}}{\partial Y_p}
  F^{\wedge p-i-1}
\wedge   \left[ * {\rm Tr} F^{\wedge p} \right] \wedge F^{\wedge i}  \right\} \wedge \delta A ~~\cr
& -  m \int_{S^{2m}}  D \left( \frac{\partial {\cal F}}{\partial Z}  {\rm Tr} F^{\wedge m-1} \right) \wedge \delta A.
\end{align}
Hence, the equation of motion derived from the pseudo-energy ${\cal E}$ is
\begin{align}
&\sum_{p=1}^m \sum_{i=0}^{p-1} (-1)^p  D \left\{  \frac{\partial {\cal F}}{\partial X_p}   \left[ F^{\wedge p-i-1} \wedge \left( * F^{\wedge p} \right) \wedge  F^{\wedge i}  \right]  \right.+ \left.  \frac{\partial {\cal F}}{\partial Y_p}
  F^{\wedge p-i-1}
\wedge   {\rm Tr}\left[ *  F^{\wedge p} \right] \wedge F^{\wedge i}  \right\} \cr
&+  m   D \left( \frac{\partial {\cal F}}{\partial Z}  {\rm Tr} F^{\wedge m-1} \right) =0~~.
\label{eqn:eom}
\end{align}
The remaining part of this article will show that the Cremmer-Scherk configuration \cite{Cremmer:1976ir} solves the above equation.
The Cremmer-Scherk configuration $A^{\rm (CS)}$ and the corresponding field strength two-form $F^{\rm (CS)}$ are, respectively,
\begin{align}
A^{\rm (CS)} &= \frac{1}{4qR^2} \zeta^a V^b \gamma_{a,b}~
~~~\mbox{and}~~~~
F^{\rm (CS)} 
=  \frac{1}{4qR^2} V^{ab} \gamma_{a,b} ~~.
\end{align}
Hereafter, the gauge field is assumed to be this Cremmer-Scherk configuration, and the notation ${\rm CS}$ is omitted.
As shown in \cite{Kihara:2011jc}, the powers of the field strength satisfy the generalized self-duality equation
\begin{align}
* F^{\wedge p} &=(4qR^2)^{m-2p}\frac{i^m (-1)^{m-p}(2p)!}{(2m-2p)!} \gamma_{2m+1} F^{\wedge (m-p)} ~.
\label{eqn:gsde}
\end{align}
The auxiliary quantities $\sigma_p$ become constant:
\begin{align}
\sigma_p &= (-1)^p (4qR^2)^{m-2p}\frac{i^m (-1)^{m-p}(2p)!}{(2m-2p)!} {\rm Tr} \gamma_{2m+1}
(4qR^2)^{-m}\frac{i^m (2m)!}{1} \gamma_{2m+1} \cr
&=  (4qR^2)^{-2p}\frac{(2p)!(2m)!}{(2m-2p)!} 2^m  ~.
\end{align}
In order to evaluate the quantities $\tau_p$ and $\rho_m$, the values of the traces of $\gamma_{ a_1,a_2,\cdots  ,a_p}$ are required.  Suppose that $k \geq 1$ and $1 \leq a_i \leq 2m, i=1,2, \cdots , 2k-1$.
${\rm Tr} \gamma_{a_1,a_2,\cdots  ,a_{2k-1} }=
{\rm Tr} \gamma_{a_1,a_2,\cdots  ,a_{2k-1} } \gamma_{2m+1}^2 = {\rm Tr} \gamma_{2m+1}\gamma_{a_1,a_2,\cdots  ,a_{2k-1} } \gamma_{2m+1}$.
Because $\{ \gamma_{2m+1} , \gamma_a \} =0$, ${\rm Tr} \gamma_{a_1,a_2,\cdots  ,a_{2k-1} }=
-{\rm Tr} \gamma_{a_1,a_2,\cdots  ,a_{2k-1} }=0$.
${\rm Tr} \gamma_{a_1,a_2,\cdots  ,a_{2k} } = {\rm Tr} \gamma_{a_1,a_2,\cdots  ,a_{2k-1} } \gamma_{a_{2k}}=  {\rm Tr}  \gamma_{a_{2k}} \gamma_{a_1,a_2,\cdots  ,a_{2k-1} }$ ($a_i \neq a_{2k}, i=1,\cdots, 2k-1$). $a_{2k} \neq a_i$ implies that $\gamma_{a_{2k}} \gamma_{a_1,a_2,\cdots  ,a_{2k-1} } =
- \gamma_{a_1,a_2,\cdots  ,a_{2k-1} }\gamma_{a_{2k}}$.
Hence, ${\rm Tr}   \gamma_{a_1,a_2,\cdots  ,a_{2k} } =0$.
The field strength two-form of the Cremmer-Scherk configuration satisfies
$F^{\wedge p} \propto V^{a_1} \wedge   \cdots \wedge V^{a_{2p}} \gamma_{a_1, \cdots ,a_{2p}}$.
Therefore, ${\rm Tr} F^{\wedge p} =0$ and
\begin{align}
\tau_p &= 0 ~, & \rho_m &=0 ~~.
\end{align}
These results show that any function of $\sigma_p, \tau_p$, and $\rho_m$ becomes constant if the gauge field is the Cremmer-Scherk configuration. 
It is thereby proven that the Cremmer-Scherk configuration solves Eq.(\ref{eqn:eom}). 
As shown above, $\displaystyle \frac{\partial {\cal F}}{\partial X_p}(\sigma_1, \cdots ,\sigma_p, \tau_1, \cdots, \tau_p, \rho_m)$, $\displaystyle \frac{\partial {\cal F}}{\partial Y_p}(\sigma_1, \cdots ,\sigma_p, \tau_1, \cdots, \tau_p, \rho_m)$, and $\displaystyle \frac{\partial {\cal F}}{\partial Z}(\sigma_1, \cdots ,\sigma_p, \tau_1, \cdots, \tau_p, \rho_m)$ become constant. In addition, ${\rm Tr}\left[  F^{\wedge p} \right]$ vanishes. Hence, the left-hand side of Eq.(\ref{eqn:eom}) reduces to
\begin{align}
(LHS)&=\sum_{p=1}^m \sum_{i=0}^{p-1} (-1)^p  \frac{\partial {\cal F}}{\partial X_p}  D \left\{    \left[ F^{\wedge p-i-1} \wedge \left( * F^{\wedge p} \right) \wedge  F^{\wedge i}  \right]    \right\}~~.
\label{eqn:redeom}
\end{align}
The generalized self-duality equation, Eq.(\ref{eqn:gsde}), implies that
\begin{align}
(LHS)&=\sum_{p=1}^m p (-1)^p  \frac{\partial {\cal F}}{\partial X_p} (4qR^2)^{m-2p}\frac{i^m (-1)^{m-p}(2p)!}{(2m-2p)!} \gamma_{2m+1}
 D F^{\wedge m-1} =0~~.
\end{align}
Hence, the Cremmer-Scherk configuration solves Eq.(\ref{eqn:eom}).

To conclude, it was shown that the Cremmer-Scherk configuration solves the equation of motion obtained from the pseudo-action defined by a function ${\cal F}(X_1, \cdots ,X_m , Y_1 , \cdots , Y_m,Z)$ of class $C^1$. 
 This class of function includes Lagrangians consisting of multi-trace terms. 
For instance ${\cal F}=X_1^3$, the Lagrangian density is 
${\cal L} = \sigma_1^3= \left( *\left[ - {\rm Tr} 
 \left\{ F \wedge * F \right\} \right] \right)^3$. 
For future work, it would be interesting to extend this case to include terms that cannot be explicitly written as functions of $\sigma_p$, $\tau_p$, and $\rho_m$, for instance, ${\rm Tr} F\wedge *(F \wedge F) \wedge *( F\wedge *(F \wedge F))$.
Classification of actions with respect to the stability condition against the Cremmer-Scherk configuration would also be an interesting problem.
The discussion in this article is applied to spaces shown in \cite{Kihara:2011jc}. 
Because the multiplications of field strength two-forms depend on the representation, 
generalization to the generic symmetric spaces is much difficult but interesting future work.


\begin{thebibliography}{99}

\bibitem{Cremmer:1976ir}
  E.~Cremmer and J.~Scherk,
  ``Spontaneous Compactification of Space in an Einstein Yang-Mills Higgs
  Model,''
  Nucl.\ Phys.\  B {\bf 108}, 409 (1976);
  ``Spontaneous Compactification of Extra Space Dimensions,''
  Nucl.\ Phys.\  B {\bf 118}, 61 (1977).


\bibitem{Tchrakian:1978sf}
  D.~H.~Tchrakian,
  ``N-Dimensional Instantons and Monopoles,''
  J.\ Math.\ Phys.\  {\bf 21}, 166 (1980);
  D.~H.~Tchrakian,
  ``Spherically Symmetric Gauge Field Configurations with Finite Action in 4
  P-Dimensions (P = Integer),''
  Phys.\ Lett.\  B {\bf 150}, 360 (1985);
  D.~O'Se and D.~H.~Tchrakian,
 ``Conformal Properties of the BPST Instantons of the Generalized Yang-Mills
  System,''
  Lett.\ Math.\ Phys.\  {\bf 13}, 211 (1987);
  Z.~Ma and D.~H.~Tchrakian,
  ``Gauge Field Systems on Cp(N),''
  J.\ Math.\ Phys.\  {\bf 31}, 1506 (1990).
\bibitem{Bais:1985ns}
  F.~A.~Bais and P.~Batenburg,
  ``Yang-Mills Duality in Higher Dimensions,''
  Nucl.\ Phys.\  B {\bf 269}, 363 (1986).
\bibitem{Kihara:2011jc}
  H.~Kihara,
  ``Generalized Self-Duality Equations of Polynomial Type in Yang-Mills
  Theories,''
	J.\ Math.\ Phys.\ {\bf 52}, 072301 (2011)
	  [arXiv:1103.0388 [hep-th]].




\end{thebibliography}
\end{document}